\documentclass[draft]{agujournal2019}
\pdfoutput=1
\usepackage{url} 
\usepackage{lineno}
\usepackage[inline]{trackchanges} 
\usepackage{soul}
\draftfalse

\journalname{JGR: Space Physics}

\begin{document}

%
%


\title{Low Pressure EUV Photochemical Experiments: Insight on the Ion-Chemistry Occurring in Titan's Atmosphere}

%
%




\authors{J. Bourgalais\affil{1}, N. Carrasco\affil{1}, L. Vettier\affil{1}, and P. Pernot\affil{2}}


\affiliation{1}{Universit\'e Versailles St-Quentin, Sorbonne Universit\'e, UPMC Univ. Paris 06, CNRS/INSU, LATMOS-IPSL, 11 boulevard d'Alembert, 78280 Guyancourt, France}
\affiliation{2}{Laboratoire de Chimie Physique, CNRS, Univ. Paris-Sud, Universit\'e Paris-Saclay, 91405, Orsay, France}




\correspondingauthor{Bourgalais}{jeremy.bourgalais@latmos.ipsl.fr}




\begin{keypoints}
\item Experiments
\item Ionosphere
\item Photochemistry
\end{keypoints}

%
%

%
%


\begin{abstract} 
{Thanks to the \textit{Cassini} spacecraft onboard instruments, it has been known that Titan's ionospheric chemistry is complex and the molecular growth is initiated through the photolysis of the most abundant species directly in the upper atmosphere. Among the pool of chemical compounds formed by the photolysis, N-bearing species are involved in the haze formation but the chemical incorporation pathways need to be better constrained.}\\ 
In this work, we performed low-pressure EUV photochemical laboratory experiments. The APSIS reactor was filled with a N$_2$/CH$_4$ (90/10\%) gas mixture relevant to the upper atmosphere of Titan. The cell was irradiated by using a EUV photon source at 73.6 nm {which has been difficult to produce in the laboratory for previous studies.}. The photoproducts (both neutral and ionic species) were monitored \textit{in situ} with a quadrupole mass spectrometer. The chemical pathways are explained by confronting experimental observations and numerical predictions of the photoproducts. \\ 
The most interesting result in this work is that methanimine was the only stable N-bearing neutral molecule detected during the experiments and it relies on N$_2^+$ production. This experimental result is in agreement with the relatively high abundance predicted by 1D-photochemical models of Titan's atmosphere and comforts methanimine as an intermediate towards the formation of complex N-bearing organic molecules.\\ 
{This experiment is only testing one part of the overall chemical scheme for Titan's upper atmosphere due to the selective wavelength but demonstrates the capability to probe the chemical pathways occurring in Titan's atmosphere by minimizing bias coming from wall surface reactions.}
\end{abstract}

%
%

\section{Introduction}
\subsection{Photochemistry of Titan's Atmosphere}
The exposition of Titan's highest layer to external sources of energy leads to the photoionization and photodissociation of the most abundant species (N$_2$ and CH$_4$). {Those processes trigger an efficient photochemistry forming small complex molecules including both neutral and ionic species. The CAPS/ELS instrument detected heavy anions up to 10,000 Da/q \cite{coates2007discovery,coates2009heavy,wellbrock2013cassini,desai2017carbon} and the IBS along with the INMS instruments measured organic positive species up to 350 Da/q. \cite{waite2007process,crary2009heavy}} \\
For most of them, photochemical models are now able to reproduce reasonably the observed abundances.\cite{horst2008origin,lavvas2008couplinga,lavvas2008couplingb,yelle2010formation,krasnopolsky2012titan,westlake2012titan,hebrard2013photochemistry,loison2019photochemical,li2015vertical,dobrijevic20161d,vuitton2019simulating} Those molecules induce a progressive formation of more complex organic compounds at lower altitudes \cite{loison2019photochemical}, which are the building block toward the formation of aerosols.\cite{lavvas2013aerosol} \\
However, in spite of all the efforts {coming from the Cassini/Huygens space mission and laboratory studies} to understand the molecular growth pathways building up complexity in the atmosphere of Titan, there are still some areas of concern with remaining unanswered questions especially regarding the formation of heavier molecules, preventing us from a complete picture of Titan's planet evolution.\cite{horst2017titan} \\
{Especially, the nitrogen incorporation in Titan's aerosols remain mostly unknown. Based on laboratory analogue studies \cite{berry2019chemical,dubois2019nitrogen}, the aerosols following mechanisms including both positive and negative ions, would be composed of nitrogen along with hydrocarbons contributing to chemical functional groups such as amines and nitriles which are of interest for exobiology. \cite{gautier2014nitrogen} \\
Moving ahead would require a deeper knowledge of the ion chemistry which} was pointed out as a significant source of neutral species \cite{vuitton2008formation,de2008coupled,krasnopolsky2009photochemical} and Plessis \textit{et al.} were the first to quantify in a new model the impact of dissociative recombination of ions on the densities of neutral species in the upper ionosphere of Titan.\cite{plessis2012production} However even recent model predictions for the density of major ions differs from the measurements of INMS of the \textit{Cassini} spacecraft up to a factor of 10 and especially at higher altitudes.\cite{vuitton2019simulating}
\subsection{EUV Irradiation Experiments} 
Since \textit{in situ} and direct observations of Titan are insufficient to understand Titan's atmosphere chemistry, the last decades have witnessed the rise of experimental laboratory simulation of Titan's atmosphere. \cite{coll2013can} The global idea is to simulate the atmosphere by mimicking the initial step, \textit{ie}. exciting a N$_2$/CH$_4$ gas mixture with different sources of energy (electrons, ions, protons, photons, X- and gamma-rays) by using plasmas or light sources, in order to produce Titan's aerosol analogues and investigate both the gas and solid phase (Titan's aerosol analogues so-called Tholins) products to understand the whole cycle. However, the most efficient processes to activate the chemistry based on N$_2$ and CH$_4$ are by interactions with solar photons. \cite{lavvas2011energy} While the less energetic (UV range) photons penetrate deeper in the atmosphere absorbing by secondary products produced by chemistry (\textit{\textit{e.g.}}, C$_2$H$_2$, C$_4$H$_2$), the high energetic photons (EUV range) are absorbed mainly by molecular nitrogen in the upper layer of the atmosphere. \\
{Previous laboratory studies investigated the gas phase products in reactors filled with relevant Titan's atmosphere gas phase mixtures using a variety of UV sources such as low-pressure Mercury and deuterium lamps limited to producing photons of 185 and 254 nm. \cite{clarke2000design,adamkovics2003photochemical,tran2008titan,yoon2014role,g2013atmospheric,sebree2014titan,cable2014identification}} However, one of the most interesting range lies above the molecular nitrogen ionization threshold (15.6 eV) leading to N$_2^+$ for which the role is still fuzzy. So far, the best way to generate a EUV photon flux is synchrotron radiation but the limited restricted access provides a limited duration for the experimental campaigns. {\cite{imanaka2007role} have shown that the formation of neutral unsaturated hydrocarbons (e.g., benzene, toluene) was enhanced when using a wavelength below 80 nm. They enounced that the increase is believed to initially come from a dissociative charge-transfer reaction between N$_2^+$ and CH$_4$, leading to the formation of unsaturated complex hydrocarbons through production of bigger ions such as C$_2$H$_5^+$ with subsequent dissociative recombination. However, they were not able to look at ions to validate this mechanism assumption.}
\subsection{Our Approach}
{In this study, we aim at corroborating the major role, highlighted by Imanaka et al., played by the photoionization of nitrogen in the formation of complex organic molecules. The neutral and ion EUV photoproducts were monitored in order to test the chemistry that is initiated by photons at a wavelength below 80 nm.}\\
As an alternative to synchrotron radiation, we used microwave plasma discharge lamps for low-pressure EUV experiments at room temperature with a Titan's relevant gas mixture (N$_2$/CH$_4$). This work is focused on the first chemical processes to constrain the N-rich chemistry supporting the chemical growth and the haze formation at high altitude in Titan's atmosphere. \\
{This work constraints the detailed chemical mechanisms by bringing for the first time information about ionic and neutral species formed in nitrogen-dominated atmospheres by EUV irradiation below the ionization threshold of nitrogen. This work is important to complete the network description and to reduce the bias in the model predictions of Titan’s ionosphere.}
\section{Experimental Method and Numerical Procedure}
\subsection{Photochemical Reactor} 
We perform a series of EUV Titan's atmosphere simulation experiments in the APSIS photochemical reactor designed to carry research on planetary atmospheres. The description of the reactor is detailed in \citeA{peng2013titan} and only a brief description is given here. \\
The reactor is a 5000 cm$^3$ stainless steel parallelepiped chamber with a length of d=50 cm. Before each experiment, the reactor is pumped down to $\sim10^{-6}$ mbar by a turbo molecular pump in order to clean out the chamber from residual gas traces. Prior to all the experiments, the photochemical reactor is also baked at 60$^\circ$C to minimize water residual traces.\\
Then the reactive gas mixture is flowed into the reactor by using a 10 sccm (standard centimeter cube per minute) range flow controller. A primary-secondary pumping unit ensures a stationary flow of reactive gas. The pressure is measured with an absolute capacitance gauge. The N$_2$/CH$_4$ (90/10\%) gas mixture is introduced continuously at a flow rate of 5 sccm, resulting in a partial pressure of P$_p$ = 3.3$\times$10$^{-3}$ mbar. The experiments are performed at a total pressure of P = 10$^{-2}$ mbar due to the injection of neon coming from the UV source (see below) and at room temperature (300 K).
  \begin{figure}
 \noindent\includegraphics[width=\textwidth]{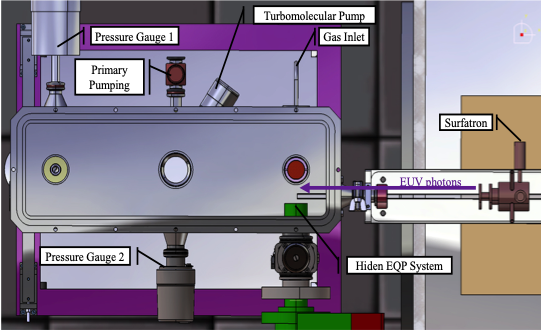}
\caption{Top view schematic diagram of the APSIS setup {with the quadrupole mass spectrometer referred as Hiden EQP System located at 1 cm from the outlet of the EUV photon source}.}
\label{APSIS}
\end{figure}
\subsection{EUV Source Coupled Windowless with the Photoreactor}
In this work we irradiate at 73.6 nm (15.6 eV) the APSIS reactor flowed with a N$_2$/CH$_4$ gas mixture. Our EUV source described in \citeA{tigrine2016microwave} and based on a surfatron discharge in a neon gas flow generates the photons emission. The flow tube used for the injection of the EUV photons is slid into the middle of the reactor. This limits the impact of the recombination of reactive species and the formation of further products, which was observed during previous laboratory simulation experiments of Titan's ionospheric chemistry at low pressure ($\sim$10$^{-5}$ mbar). \cite{thissen2009laboratory} The source is coupled windowless with our photoreactor. The distance between the end of the discharge and the entrance of the reactor is optimized to ensure that electrons and neon metastable atoms are quenched before arriving to the reactive zone: EUV photons are the only source of energy irradiating the N$_2$/CH$_4$ gas flow. The surfatron is operated with a power of 70 W, a frequency of 2.6 GHz and a pressure of 1.5 mbar, ensuring a photon flux of $\sim$10$^{14}$ ph s$^{-1}$ cm$^{-2}$. Considering the partial pressure of N$_2$ of 3.3$\times$10$^{-3}$ mbar, the absorption cross-section of N$_2$ at 73.6 nm and the length of the reactor (d = 50 cm), the optical depth is 0.27 leading to a transmission of 76\%.
\subsection{Comparison of the Experimental Conditions with Titan's Ionosphere}
In Titan's ionosphere from about 900-1500 km, the methane abundance is not constant. It increases from 2\% up to 10\% due to molecular diffusion of lighter species. \cite{waite2005ion} Thus a N$_2$/CH$_4$ (90/10\%) gas mixture has been chosen to improve the detection of photoproducts in our reactor within the representative concentration range of methane in Titan's ionosphere. \\
Based on \citeA{westlake2011titan,snowden2013thermal,snowden2014thermal} on Huygens Atmospheric Structure Instrument (HASI) the temperature range in Titan's ionosphere goes from 150 to 200 K. We are working at room temperature. However the ion-molecule reactions, which are the main processes in the chemical regime simulated in this experiment, show a low temperature dependence. \cite{dutuit2013critical} Thus there is no significant stake at working in Titan's temperature range. \\
We work at a lower pressure than almost all the other experimental simulations mimicking Titan's ionospheric chemistry (except the experiment by \citeA{thissen2009laboratory}), but it is still a higher pressure than in Titan's ionosphere. The limitation is experimental to avoid wall reactions. At 0.01 mbar, the pressure is low enough to minimize termolecular effects as in Titan's ionosphere, so that the higher pressure is expected to enhance the kinetics without affecting the general chemical scheme.\\
Our EUV source provides monochromatic photons at 73.6 nm, and no large EUV spectral range as in the solar spectrum. The experiment enables to explore ionization regimes of both N$_2$ and CH$_4$ as in Titan ionosphere, but mostly neglects contributions of neutral photoproducts from N$_2$ and CH$_4$. 
\subsection{Mass Spectrometry Analysis}
The tip of the QMS is set up 1 cm perpendicularly to the axis of the photon flux after the exit quartz flow tube. Gas sampling is done through a small pinhole of 100 microns in diameter (see Figure \ref{APSIS} for a top view schematic diagram of the APSIS setup). This experimental setup allows for in situ measurements of the gas phase composition by using a Hiden EQP System. This system is a high sensitivity ($<$ 1 ppm) quadrupole mass spectrometer combined with an energy analyzer. It monitors and characterizes positive and negative ions, and reactive neutral species. \\
Neutral species are ionized in the electron ionization chamber, settled with an electron acceleration voltage of 70V. The detector has a mass resolving power of 1 amu and covers the 1-200 amu range. The signal expressed in count-per-second (cps) is obtained over an integration period called dwell time and set to 100 ms. A pressure of 10$^{-8}$ mbar in the mass spectrometer is ensured by a turbo molecular pump, avoiding secondary reactions during the transfer of the gas phase through the ion optics of the mass spectrometer. \\
In the following experimental mass spectra are the results of 10 scans average over the 200 amu range and the uncertainty of the signal are given by the standard deviation of the 10 scans for each mass.\\
Additional experiments have been performed by changing the dwell time and no significant difference is observed. In all the spectra presented below, counts $<$10 cps correspond to background noise. Thus, the presented mass spectra thereafter show the direct signal registered without background subtraction. 
\subsection{0D-Photochemical Model}
To interpret the experimental mass spectra , a 0D-photochemical model developed by \citeA{peng2014modeling} is used to reproduce the chemistry obtained from the gas-phase in the APSIS reactor.  The model includes 244 species (125 neutrals and 119 ions) interacting with each other through 1708 reactions (33 photolysis, 415 bimolecular reactions, 82 termolecular reactions, 574 dissociative recombinations, 604 ion–molecule reactions) and takes into account the associated uncertainties. \cite{hebrard2006photochemical,hebrard2009measurements,plessis2010knowledge,carrasco2008sensitivity}. {Oxygen chemistry is included, since H$_2$O is a trace gas in the reactor.} This model implements a consistent treatment of the separation between photolysis cross-section and branching ratios and it is an ion–neutral coupled model imvolving a state-of-the-art dissociative recombination scheme. \cite{plessis2010knowledge} \\
The model is run in the same conditions as the experiments (pressure, wavelength, photon flux, geometry of the reactor) and is integrated until the temporal profile of all the masses reaches a stationary state, ie. 0.01 s. Tests have been performed at a longer reaction times and no difference is observed in the photoproducts distribution at this pressure. Then the predicted stationary state mole fractions of the photoproducts are compared to the experimental data. The comparison enables to validate the model.\\
Then, a complementary Monte Carlo simulation (500 runs) is launched, for uncertainty and sensitivity analysis. These calculations are used to identify the key reactions and the dominant growth pathways in the photochemical reactor. \cite{hebrard2009measurements}
\section{Results and Discussion}
\subsection{Ions Measurements and their Numerical Simulation}
  \begin{figure}
 \noindent\includegraphics[width=\textwidth]{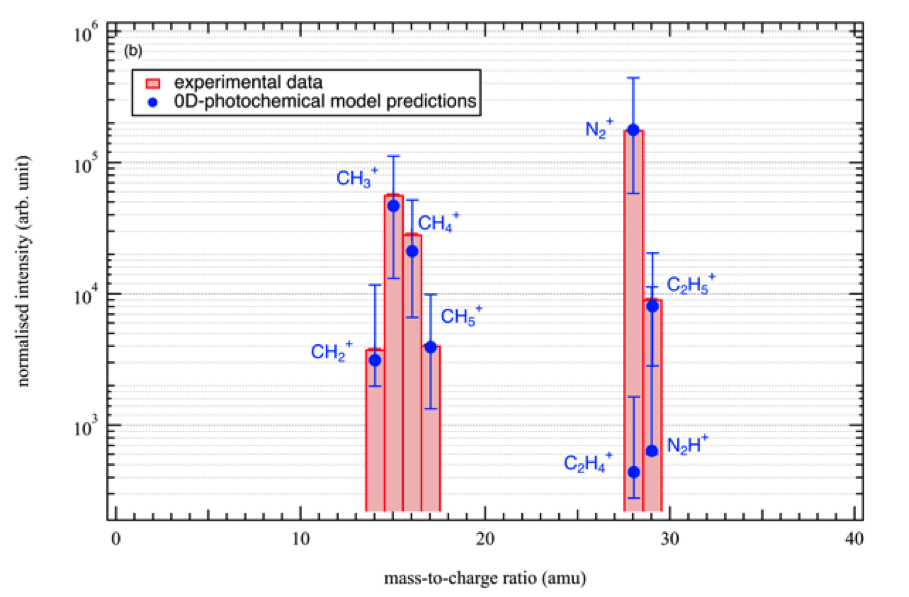}
\caption{\textit{In situ} gas-phase mass spectrum of ionic species in an APSIS experiment with a N$_2$/CH$_4$ (90/10 \%) gas mixture (red bars) compared to simulated relative abundances of the reference case given by the 0D-photochemical model (blue points). Standard deviation ($\sigma$ in 10 measurements of each mass is displayed). Experimental and numerical data are scaled to N$_2^+$.}
\label{ions}
\end{figure}
The simulated relative abundances of all the predicted ions are displayed in Figure \ref{ions}. A full agreement is found between the simulated predictions and the APSIS mass spectrum measurements. m/z 14, 15, 16, 17 are attributed to CH$_2^+$, CH$_3^+$, CH$_4^+$, CH$_5^+$ ions. N$_2^+$ is the major ion at m/z 28. C$_2$H$_4^+$ is also predicted by the model but is hidden experimentally by the higher contribution of N$_2^+$. m/z 29 is a combination of C$_2$H$_5^+$ and N$_2$H$^+$, with C$_2$H$_5^+$ as major ion considering the predicted densities from the photochemical model. In our experience, the ion content is mainly composed of primary ions N$_2^+$, CH$_4^+$ and CH$_3^+$, but a few secondary ions CH$_5^+$, C$_2$H$_5^+$ and N$_2$H$^+$ are also present.\\
The lack of peaks at m/z 18 and 19 is also a result with our experiment. Given the high proton affinity of ammonia and water, the absence of their associated protonated ions NH$_4^+$ (m/z 18) and H$_3$O$^+$ (m/z 19) in the ion mass spectra confirm the low abundance of these both stable molecules in the reactor. The absence of water is a good indicator of the very low contamination level. The absence of ammonia is an interesting diagnosis of the absence of wall-effects in our experiment, as the formation of ammonia is known to be mainly driven by wall effects. \cite{touvelle1987plasma,thissen2009laboratory}
\subsection{Comparison to other Studies Probing the Ion Content in Titan-like Photochemical Experiments}
To our knowledge only two previous studies measured positive ions in N$_2$/CH$_4$ gas mixtures irradiated by far-UV photons to address experimentally the complexity of the mass spectra collected by the INMS instrument in Titan's ionosphere. \cite{berry2019chemical,thissen2009laboratory}\\
Our work is very complementary to the approach chosen by \citeA{berry2019chemical}. They chose to produce efficiently large C$_x$H$_y$N$_z^+$ ions to identify possible large ions participating to the formation of Titan's haze at high altitudes. For that purpose, they irradiated N$_2$/CH$_4$ gas mixtures at almost atmospheric pressure (840 mbar) with a Deuterium UV source. The high pressure favors termolecular reactions and the photon wavelengths were larger than 115 nm, so unable to directly photoionize CH$_4$ and N$_2$. The result is that they produce large ions (up to m/z 400) with formation pathways, which are not representative of Titan's direct ionization processes. In our case we reproduce as faithfully as possible the first ionizing reaction processes coupling N$_2$ and CH$_4$, but our experiment is limited to the simulation of the primary and a few secondary ions from the whole chemical network occurring in Titan's ionosphere. \\
The results by \citeA{thissen2009laboratory} are more directly comparable to ours. They also worked at low pressure with pulsed gaz injections leading to a pressure varying in their reaction cell between 10$^{-6}$ and 5$\times$10$^{-4}$ mbar. Then they irradiated the N$_2$/CH$_4$ gas flow with ionizing EUV radiations with photon wavelengths varying between 91.8 nm (13.5 eV) and 46.8 nm (26.5 eV). The main ions observed in this case are at m/z 18, 19, 27, 28 and 29, attributed respectively to NH$_4^+$, H$_3$O$^+$, HCNH$^+$, C$_2$H$_4^+$, and C$_2$H$_5^+$. They observe an important increase of the total ion density in their cell at 15.6 eV, the ionization threshold of N$_2$, reinforcing experimentally the predictions of \citeA{imanaka2007role} that N$_2^+$ has a major role in the ion chemistry coupling N$_2$ and CH$_4$. H$_3$O$^+$ is unfortunately the major ion in their experiment, revealing an important relative contamination by residual water in the reactive cell. NH$_4^+$ and HCNH$^+$ are also found to be mainly explained by wall effects, showing that the sizing of their reactor cell is a strong limitation to study gas-phase ion chemistry as occurring in Titan's ionosphere. In spite of these experimental difficulties, they observe that C$_2$H$_5^+$ is one of the main secondary ions produced by the first ion-molecule reactions triggered by photoionization of N$_2$ and CH$_4$. Our work is in agreement with this finding as supported by our experimental mass spectrum and simulated mole fraction (Figure \ref{time_profile_MC}). Without wall effects and water contamination, we moreover observe clearly the primary ions and the formation of CH$_5^+$. \\
None of these  experiments enables to probe large molecular growth processes, as the short residence time of the species in the reactors prevents the formation of key neutrals C$_2$H$_2$ and C$_2$H$_4$ for further the ion chemical growth. \cite{peng2014modeling,westlake2014role} A slightly further chemical growth with the additional detection of C$_3$H$_3^+$, C$_3$H$_5^+$ and C$_4$H$_5^+$ was actually observed in \citeA{thissen2009laboratory} when adding C$_2$H$_2$ and C$_2$H$_4$ in their initial gas mixture. 
\subsection{Comparison with INMS Ion Measurements in Titan's Ionosphere}
The INMS instrument detected positive ions in Titan's ionosphere with m/z up to 100. The ion-molecule chemical growth is not as efficient in our experiment, with the largest positive ion detected at m/z 30, and in spite of the higher EUV photon flux used in our experiment (10$^{14}$ ph s$^{-1}$ cm$^{-2}$). The reason for this discrepancy is the dynamical transport, which differs strongly in Titan's ionosphere (molecular diffusion) and in our reactor (uniform advective velocity). \\
To ensure a low pressure in the reactor, a primary-secondary pumping unit extracts the gas continuously. An equilibrium in pressure is found with the continuous inflow rate. The residence time t$_r$ of the gases in our reactor is 0.1 s, with the formula:  t$_r$ = (P$_{tot}$/P$_0$)$\times$(V/D$_v$), with P$_{tot}$ the pressure in the reactor, P$_0$ is the reference standard pressure of 1 bar, V the volume of the reactor (in cm$^3$) and D$_v$ the gas inflow (in cm$^3$ min$^{-1}$).\\
In Titan's ionosphere, the molecular diffusion coefficient D of the mixture N$_2$/CH$_4$ is about 10$^8$ cm$^2$ s$^{-1}$ at 1000 km of altitude. \cite{toublanc1995photochemical} Using a typical transport corresponding to the scale height H on Titan, ie. 20 km, we obtain a rough estimation of the timescale of the molecules in Titan of 10$^4$ s with the relation t$_r$$\sim$H$^2$/D.\\
The dynamical transport is therefore 10$^5$ faster in our experiment compared to Titan's ionosphere, explaining why we are able to simulate the first steps of the N$_2$/CH$_4$ ion chemical network (primary photoionization and first ion-molecules reactions), but the present experimental conditions cannot ensure further ion-molecule reactions to observe the larger ions seen in Titan's ionosphere. Several experimental strategies can be considered to explore further the ion-molecule chemical network. \citeA{thissen2009laboratory} for example proposed to enrich the initial gas mixture with a few key neutrals (C$_2$H$_2$ and C$_2$H$_4$) known to be involved in ion chemical growth. \citeA{berry2019chemical} proposed a different approach by using assumed far from representative experimental conditions (high pressure and irradiation with a deuterium lamp) but enabling to rapidly reach the production of larger C$_x$H$_y$N$_z$ ions.  \\
We can now compare the light ions observed in Titan's ionosphere and in our experiment (Table 1). Methane primary ions are visible in both spectra: CH$_3^+$ and CH$_4^+$. The secondary ions CH$_5^+$ and C$_2$H$_5^+$ are also in agreement in both spectra, confirming the relevance of our experiment to simulate the first step of the N$_2$/CH$_4$ ion chemistry. N$^+$ is observed in Titan's ionosphere and not in our experiment because the 73.6 nm wavelength for the photons generated with our EUV source is below the dissociative ionization threshold of N$_2$. The lack of N$^+$ to participate to the rest of the chemical scheme has no consequence, as the main nitrogen ion driving ion-neutral chemical reactivity is N$_2^+$. \cite{imanaka2007role} Some C$_2$ carbocations, C$_2$H$_2^+$, C$_2$H$_3^+$, HCNH$^+$, CH$_2$NH$_2^+$, and C$_2$H$_7^+$ are not observed in our case due to the too fast transport in our case, preventing successive bimolecular reactions to occur.\\
The last difference is the observation of protonated ammonia NH$_4^+$ in INMS spectra and not in our case. A debate exists on the density of ammonia in Titan's ionosphere. Most photochemical models fail in reproducing density of ammonia high enough to explain such a high ion density of NH$_4^+$. \cite{dobrijevic20161d,vuitton2019simulating} A high density of ammonia is also in disagreement with the density measured in the stratosphere. \cite{nixon2013detection} {In the atmosphere of Titan, \citeA{yelle2010formation} suggested that ammonia is produced primarily from amino radical (NH$_2$) reaction with methylene-amidogen radical (H$_2$CN). H$_2$CN is mainly formed through the ground state nitrogen atom N($^4$S) reaction with methyl radical (CH$_3$) and constitutes the main exit channel of the reaction. \cite{dutuit2013critical} Thus, the ammonia and its protonated form NH$_4^+$ found in the most recent model \cite{dobrijevic20161d,vuitton2019simulating} put into question the validity of the INMS ammonia detection, with potential artifact due to wall effects and/hydrazine fuel from the \textit{Cassini} space probe \cite{cui2009analysis,magee2009inms}.\\
In this work, the comparison with INMS is only mildly relevant in this case because of the limited energy input for the experiment compared to the energy input to Titan's atmosphere. The absence of both ammonia and its protonated form in this work could be explained by the nitrogen photolysis leading mainly to the formation of the excited state N($^2$D) and N$_2^+$. Thus, the formation of ground state nitrogen atom N($^4$S) is not conducived, preventing us to address the chemical pathways of the formation of ammonia for instance \citeA{yelle2010formation}. However, this work has demonstrated the effectiveness of probing the chemical pathways that result from the formation of N$_2^+$ in Titan's atmosphere by minimizing biases from wall reactions.}

 \begin{table}
 \caption{Comparison of light ions (m/z $<$ 35) detected by the INMS instrument in Titan's ionosphere \cite{vuitton2006nitrogen} and this work}
 \centering
 \begin{tabular}{c c c}
 \hline
  m/z  & INMS & This Work  \\
 \hline
   14  &  N$^+$ & - \\
   15  & CH$_3^+$  &CH$_3^+$ \\
   16  & CH$_4^+$ & CH$_4^+$ \\
   17  & CH$_5^+$ & CH$_5^+$ \\
   18  & NH$_4^+$ &- \\
   26  & C$_2$H$_2^+$ & - \\
   27  & C$_2$H$_3^+$ & - \\
   28  & HCNH$^+$ & N$_2^+$  \\
   29  & C$_2$H$_5^+$ &  C$_2$H$_5^+$  \\
   30  & CH$_2$NH$_2^+$ & - \\
   31  & C$_2$H$_7^+$ & - \\
 \hline
 \multicolumn{2}{l}{}
 \end{tabular}
 \end{table}

 \subsection{Anions Measurements}
 
{Despite many attempts, anions were not detected in this work and the photochemical model does not include the formation of negative species. In this work, the main source of energy are the EUV photons and the anions are mostly formed by dissociative attachment of photoelectrons in our reactor onto neutral species. Thus, the limit of sensitivity is too high to detect the concentration of negative species.\\ Additionally, the selectivity of the wavelength, 73.6 nm, in this work does not conducive the formation of anions. For instance, based on the first anionic chemistry implemented in an ionospheric model \cite{vuitton2009negative} the major production reactions of the main negative specie CN$^-$ are proton transfers from HCN by small anions such as H$^-$ but the formation of HCN is favored using wavelength higher than 80 nm, thus inhibited in this work.}

\subsection{Neutral Measurements}

 \begin{figure}
 \noindent\includegraphics[width=\textwidth]{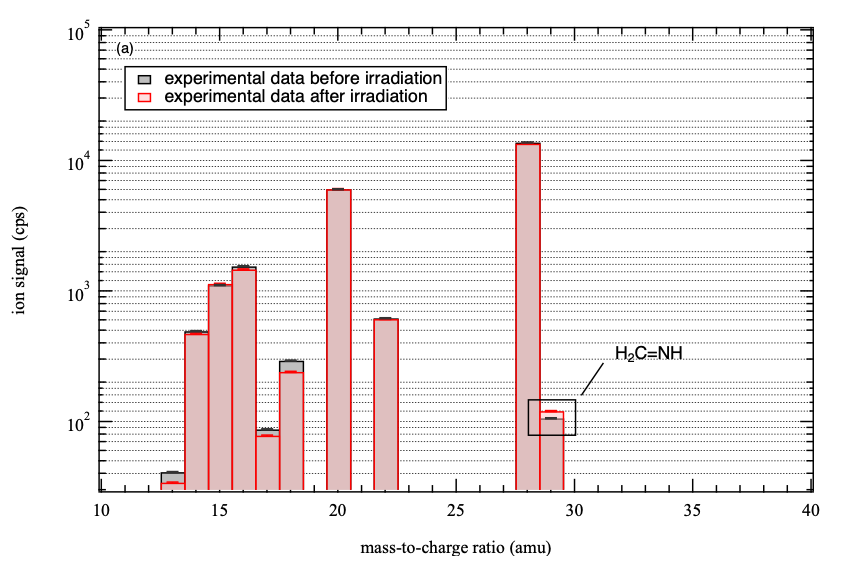}
\caption{\textit{In situ} non-normalized gas-phase mass spectrum of neutral species in an APSIS experiment with a N2/CH4 (90/10 \%) gas mixture before (grey bars) and after (red bars) EUV irradiation. Standard deviation ($\sigma$ in 10 measurements of each mass is displayed).}
\label{neutre_total}
\end{figure}

As discussed previously, the rapid transport of the molecules in our reactor prevents to observe large neutrals and/or successive bimolecular reactions. The neutral mass spectra are moreover complicated by the dominant signature of the reactants N$_2$ and CH$_4$. These two species contribute actually to numerous ion peaks, both through complex fragmentation pattern by electronic impact at 70 eV in the ionization chamber of the mass spectrometer and through their isotopic natural contributions. The number of neutral photoproducts that we observe is therefore limited (Figure \ref{neutre_total}).\\
{Figure \ref{neutre_total} shows an \textit{in situ} non-normalized gas-phase mass spectra of neutral species in their stationary point in an APSIS flow reactor experiment before and after 15 min EUV irradiation at a total pressure of 10$^{-2}$ mbar. The species are detected at m/z 12, 13, 14, 15, 16, 17, 18, 20, 22, 28 and 29. m/z 16 is attributed to methane CH$_4$ and m/z 12, 13, 14 and 15 to CH$_4$-electron ionization fragments C, CH, CH$_2$, CH$_3$ respectively. m/z 17 and 18 are attributed to OH and H$_2$O respectively. Due to the high sensitivity of the mass spectrometer H$_2$O is detected as impurity and OH is coming from the fragmentation of H$_2$O due to the electron ionization at 70 eV used by the mass spectrometer. This contribution is coming from both residual air in the mass spectrometer and in the photochemical reactor despite an efficient baking before the experiment due to the walls of a steel bellow surrounding the transfer line. {However, its presence has no impact on the chemistry at the wavelength used in this work because the dominant exit route during water photolysis is the formation of H$_2$O$^+$ and H$_3$O$^+$ ions as shown in \citeA{thissen2009laboratory}. However, we do not see them in our ion spectra (cf. Figure \ref{ions}) and the contribution of H$_2$O is included in our 0D-photochemical model to confirm that no O-species are formed at m/z 29.}
{m/z 20 and 22 are attributed to neon and its first isotope $^{22}$Ne respectively. Both peaks intensity does not change when the surfatron is on since the neon is used only to produce the EUV photons and does not interact with the chemistry in the reactor}.}

 \begin{figure}
 \noindent\includegraphics[width=\textwidth]{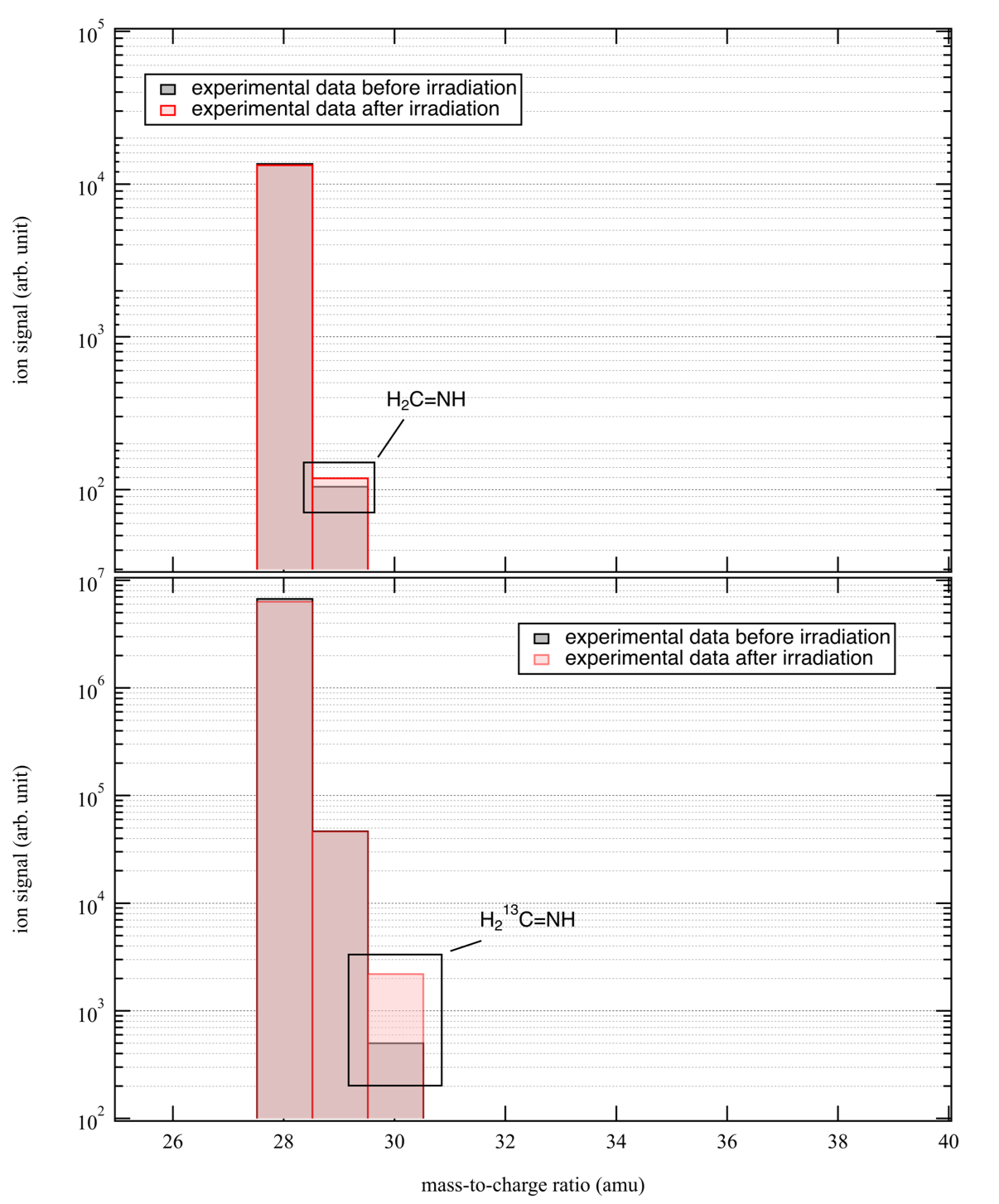}
\caption{(Upper panel) \textit{In situ} non-normalized gas-phase mass spectrum of neutral species in an APSIS experiment with a N$_2$/CH$_4$ (90/10 \%) gas mixture before (grey bars) and after (red bars) EUV irradiation. Standard deviation ($\sigma$ in 10 measurements of each mass is displayed). (Lower panel) In situ non-normalized gas-phase mass spectrum of neutral species in an APSIS experiment with a N$_2$/$^{13}$CH$_4$ (90/10 \%) gas mixture before (grey bars) and after (red bars) EUV irradiation. {Mass spectra has been cut off in mass channels for the sake of clarity}.}
\label{neutre}
\end{figure}

The only significant production that is detectable in our experiment is at m/z 29 (Figure \ref{neutre}, upper panel). Four attributions are possible for this mass signature: methanimine CH$_2$NH, N$_2$H or the fragment C$_2$H$_5^+$ coming from the ionization of ethane (C$_2$H$_6$), and HCO. N$_2$H can easily be ruled out. It would require an unlikely radical-radical reaction between NH and NH$_2$. \cite{hebrard2009measurements} The absence of ethane on the mass spectrum at m/z 30 rules out the contribution of C$_2$H$_5^+$. The oxygenated species HCO is also unlikely, since the low pressure in our experiment minimizes any bias coming from contamination. The most likely product is therefore methanimine at m/z 29. Additional experiments were performed with isotopic labeled gas mixture ($^{13}$CH$_4$) to disambiguate our identification of methanimine. Figure \ref{neutre} (lower panel) displays a mass spectrum resulting from the accumulation of 80 scans due to the low intensity of neutral species. A signal at m/z 30 clearly appeared after the irradiation, matching with a $^{13}$CH$_2$NH compound whereas N$_2$H and $^{13}$C$_2$H$_5$ would appear at m/z 29 and m/z 31 respectively. The isotopic experiment confirms that the main product observe at m/z 29 in the regular experiment (without isotopic labeling) is methanimine.
\subsection{Comparison of the Neutrals Detection with the 0D-photochemical Model Prediction}
\begin{figure}
 \noindent\includegraphics[width=100mm]{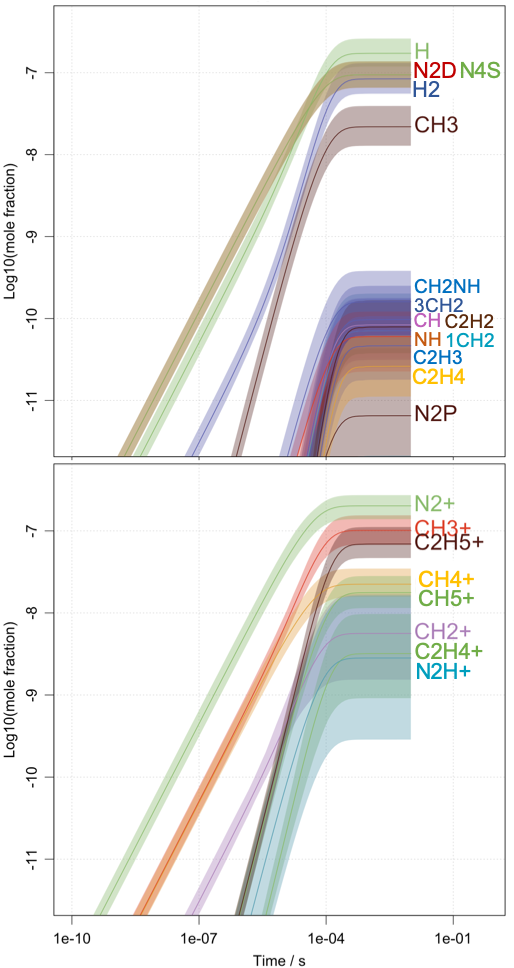}
\caption{{Simulated evolution of the mole fraction compared to N$_2$ of the main neutral (upper panel) and ionic photo-products (lower panel) until the stationary state, obtained from 500 runs of the 0D-photochemical model. All species with a mole fraction above 10$^{-11}$ are represented}.}
\label{time_profile_MC}
\end{figure}

Based on the temporal profile predictions of the 0D-photochemical model for neutral species displayed in Figure \ref{time_profile_MC}, the most abundant neutral molecules besides atoms (H and N) are methyl radical CH$_3$ and methanimine H$_2$C=NH. The other predicted molecules are CH, CH$_2$ and NH. The formation of the small species, H, H$_2$, N, NH, CH, CH$_2$ and CH$_3$ are not detectable in our experimental neutral mass spectra as their contribution is hidden by the ion fragments (and isotopes) of N$_2$ and CH$_4$. But methanimine is actually the main ''heavy'' neutral (with 2 heavy atoms), predicted by the model, in agreement with our experimental observation. 
\subsection{Sensitivity Analysis and Reaction Scheme}
\begin{figure}
 \noindent\includegraphics[width=\textwidth]{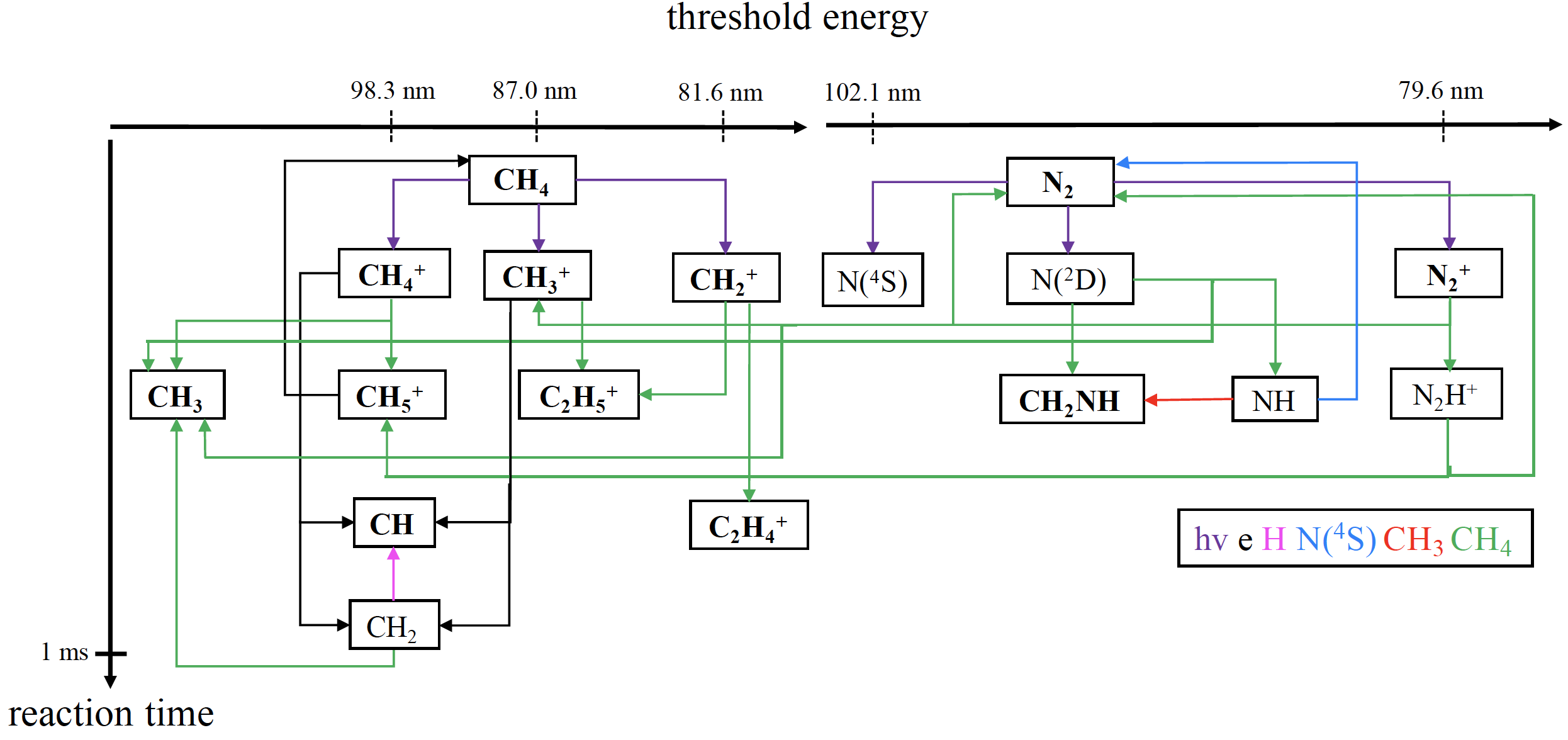}
\caption{Chemical flowchart representing the most important reactions taking place in the APSIS reactor, according to the 0D-photochemical model. Bold chemical compounds are the detected species in the experiment.}
\label{flowchart}
\end{figure}

To explain the formation of the ions and of methanimine detected in this work we perform Monte Carlo simulations for uncertainty and sensitivity analysis to identify the key, formation and loss, reactions. The main reactions involving the ions detected experimentally are displayed in Figure \ref{flowchart}. \\
At the energy of 15.6 eV and the overall pressure (10$^{-2}$ mbar) of the experiments, CH$_4$ is ionized and for one part dissociated to form the dominant ions CH$_3^+$ and CH$_4^+$ and N$_2$ is ionized to give N$_2^+$. These primary ions subsequently react with methane to produce secondary ions. CH$_3^+$ and CH$_4^+$ react with CH$_4$ to produce C$_2$H$_5^+$ and CH$_5^+$ respectively. N$_2^+$ also reacts with methane. It leads mainly through the dissociative charge-transfer reaction to CH$_3^+$, but a second channel also gives way to the formation of N$_2$H$^+$. \\
Finally the most abundant neutral molecule detected and the first N-bearing molecule in our experimental conditions is methanimine H$_2$C=NH which is formed through mainly the N($^2$D)+CH$_4$ reaction. A small proportion could additionally come from the NH+CH$_3$ reaction. \cite{rosi2018formation} The detection of methanimine is in agreement with previous kinetic modeling studies where H$_2$C=NH was mentioned to probably form through the first electronically excited state of atomic nitrogen, N($^2$D) reacting with methane. \cite{balucani2010formation,yelle2010formation}
\subsection{Comparison of the Neutrals Production with Neutrals in Titan's Ionosphere}
We notice in neutral and ionic data the absence of ubiquitous Titan's atmosphere species such as acetylene (C$_2$H$_2$) and ethene (C$_2$H$_4$), in both numerical predictions and experimental observations. This absence is due to the short residence time of the gas phase in the photoreactor, but also to the specific wavelength used in this experiment (73.6 nm) which leads mainly to the formation of the excited state N($^2$D) and N$_2^+$ from nitrogen photolysis. \cite{dobrijevic20161d} The formation of HCN is also unlikely in our experiment: it requires the presence of N($^4$S) which is a minor channel in the dissociation of N$_2$ at the wavelength employed in our experiment and where N($^2$D) is assumed to be dominant. \cite{lavvas2011energy,dobrijevic20161d} Protonated methanimine CH$_2$NH$_2^+$ is also not observed in this configuration but the 0D-photochemical model predicts a density of the species under the sensitivity level of the instrument. \\
The non-detection of small neutral hydrocarbons (C$_2$H$_2$, C$_2$H$_4$) is corroborated by the predictions of the photochemical models. However \citeA{tigrine2016microwave} suspected the formation of HCN and C$_2$N$_2$, two major Titan compounds, using the same experimental setup but at a higher pressure (1 mbar) where the chemical regime was different. They were not able to address the first products of the chemistry due to the use of a less sensitive mass spectrometer connected to the chamber through a capillary tube, but their results combined with this study show the ability of our approach to explore the chemistry of nitrogen species formed in a Titan atmospheric context.
\subsection{Importance of Methanimine} 
The chemistry of small unsaturated species with nitrogen functional groups such as imine (R-CH=N-R') are of interest in exobiology as nitrogen sources in organic molecules to form prebiotic molecules since they have been observed in many astrophysical environments \cite{lovas2006detection,loomis2013detection,zaleski2013detection} and planetary atmospheres. \cite{vuitton2006nitrogen}
The simplest imine, methanimine (H$_2$C=NH), is a highly reactive closed-shell molecule due to its C=N double bond thus considered as an abiotic precursor of amino acids and nucleobases. \cite{balucani2012elementary} It might also be important for the formation of aerosols that make up Titan's haze since Tholin's (Titan analogue aerosols formed in laboratory) structure were found to be rich in nitrogen. \cite{gautier2014nitrogen,trainer2012nitrogen,he2012nmr} Methanimine is also thought as a good candidate for the formation of nitrogen organic compounds in interstellar ice analogs via polymerization. \cite{skouteris2015dimerization} H$_2$C=NH has been defined as a gas-phase interstellar molecule since the 1970's \cite{godfrey1973discovery} and observed in many interstellar environments such as molecular clouds \cite{qin2010high}, circumstellar envelopes \cite{tenenbaum2010comparative} and starburst galaxies. \cite{salter2008arecibo} Finally, Methanimine was detected indirectly in Titan's thermosphere from the analysis of the ion spectra recorded by the Ion Neutral Mass Spectrometer (INMS) of the \textit{Cassini} spacecraft with the detection of the protonated methanimine CH$_2$NH$_2^+$. Inferred from this ion concentration a relatively high mole fraction of 10$^{-5}$ for neutral H$_2$C=NH was estimated around 1100 km in the atmosphere. \cite{vuitton2006nitrogen,vuitton2007ion} Despite this predicted relatively high abundance this compound is not taken yet into account in \textit{Cassini}-INMS neutral mass spectra analysis. \cite{cui2009analysis,waite2005ion} \\
Recent photochemical models of Titan's ionosphere\cite{lavvas2008couplinga,lavvas2008couplingb,loison2015neutral,dobrijevic20161d,vuitton2019simulating} failed at reproducing the quantity of methanimine inferred from the amount of protonated methanimine measured by the INMS and overestimated its mole fraction. It should be noted that this disagreement could come from the contribution of different species at m/z 30 in addition to protonated methanimine, even if the resolution of the INMS makes it impossible to conclude. However, the models corroborated the large abundance of H2C=NH in the upper atmosphere due the coupling of ion and neutral chemistry. \cite{yelle2010formation} Nevertheless, the discrepancy between models and observations in the upper atmosphere for such a simple compound sheds light on a missing chemistry. Especially, the ion chemistry was pointed out as a significant source of neutral species \cite{vuitton2008formation,de2008coupled,krasnopolsky2009photochemical,yelle2010formation} and Plessis \textit{et al.} were the first to quantify in a new model the impact of dissociative recombination of ions on the densities of neutral species in the upper ionosphere of Titan. \cite{plessis2012production}  However even recent models predictions for the density of major ions differs from the measurements of INMS up to a factor of 10 and especially at higher altitudes. \cite{vuitton2019simulating}\\
Under the physical conditions (temperature and pressure) of Titan's upper atmosphere, H$_2$C=NH can react rapidly with radical or ionic species since most reactions involving ionized species are barrierless. According to chemical kinetics modeling studies, the pathway leading to the formation of methanimine is thought to be driven mainly by two reactions: NH+CH$_3$ \cite{redondo2006theoretical} and N($^2$D)+CH$_4$. \cite{casavecchia2002crossed,balucani2009combined,balucani2010formation} H$_2$C=NH may be also recycled through electron recombination of CH$_2$NH$_2^+$ but with a limited effect. \cite{yelle2010formation} Generally, reactions involving neutral species, that are under low temperature conditions in planetary atmospheres, are not efficient processes since they need to go through high energy barriers. However, transient species, such as the first electronically excited state of atomic nitrogen, N($^2$D), bring enough energy to overcome these barriers. Furthermore, N($^2$D) reactions with small hydrocarbons are suspected to be the source of the first N-containing organic molecules such as NH. \cite{rosi2018formation} Methanimine's high proton affinity \cite{hunter1998evaluated} allows to consider the ionic chemistry as the main sink for H$_2$C=NH by a proton abstraction from most of the protonated species, such as HCNH$^+$.\\ 
Even if methanimine is estimated to be relatively abundant, it is difficult to investigate it in Titan simulation laboratory experiments because it is a transient species and even if it is very sensitive to the UV light, \cite{teslja2004electronic} the absorption of methanimine is not known down to Lyman-alpha line wavelength (121.567 nm). However, H2C=NH has recently been reported in UV photochemical reactor experiments \cite{he2018gas} and was predicted by a 0D-photochemical model dedicated to the interpretation of the chemistry occurring in such setups. \cite{peng2014modeling} \\
\citeA{berry2019chemical} used a very high resolution Time-of-Flight Mass Spectrometer with an Atmospheric Pressure interface (APi-ToF) to measure the ions during photochemical reactions between methane and nitrogen above Ly-$\alpha$ and they measured a species with the formula CH$_4$N$^+$ matching the potential formation of methanimine. By contrast, methanimine has been detected in early plasma experiments \cite{carrasco2012volatile,horst2018laboratory} and mentioned as a characteristic product since it requires the presence of the first electronically excited state of atomic nitrogen to be formed.\\
Our study validates experimentally the formation of neutral methanimine in low pressure EUV experiments at room temperature with a Titan's relevant gas mixture (N$_2$/CH$_4$). 
\section{Conclusions}
In this work we irradiated a N$_2$/CH$_4$ (90/10 \%) gas mixture at a gas pressure of P = 10$^{-2}$ mbar at room temperature with EUV photons at 73.6 nm. The photoproduct neutral and ionic species were monitored \textit{in situ} using a Hiden EQP System. Then the experimental ion data were compared to the prediction abundances of a 0D-photochemical model allowing a qualitative comparison between predicted and observed neutral species. Monte Carlo based sensitivity analysis allowed us to identify the key formation and loss reactions.\\ 
Regarding the ion product distribution, our results are in agreement with previous synchrotron experiments. Those mentioned an increase in the formation of hydrocarbon species over 15.6 eV, due to the ionization threshold of molecular nitrogen. \cite{imanaka2009euv,thissen2009laboratory} In our experimental conditions, N$_2^+$ plays the role of an activator/trigger by doping the formation of CH$_3^+$ in the production of neutral and positively charged hydrocarbons. Dissociative charge-transfer reaction of CH$_4$ with primary ions followed by subsequent recombination with electrons might play an important role in the formation of the first neutral hydrocarbons. \cite{geppert2004first} Those unsaturated species are believed to open the road towards the formation of heavier species such as aromatic compounds. \cite{wilson2003mechanisms,moses2005photochemistry,waite2007process} Imanaka \& Smith also showed that a 60nm irradiation of N2/CH4 mixture leads to N-rich aerosols through nitrogen intermediate species in the gas phase. \cite{imanaka2010formation}\\
Our study suggests that methanimine could be one of these key compounds explaining N-enrichment of aerosols in Titan's atmosphere. Our conjugated approach coupling numerical simulation with experimental measurements of ionic and neutral species allowed also to clearly identify the formation of methanimine H$_2$C=NH as the main N-bearing molecule in our experiment. We found that it was in larger abundance than any other neutral stable hydrocarbons, due to the monochromatic irradiation at 73.6 nm. This finding is in agreement with Titan modeling studies in the literature where the high abundance of methanimine is driven by N($^2$D) + CH$_4$. \cite{yelle2010formation} However, this species was not taken into account in previous \textit{Cassini}-INMS neutral mass spectra analysis due to the lack of a cracking pattern of H$_2$C=NH. \cite{waite2005ion,cui2009analysis} This result suggests to consider future data treatment involving this additional important nitrogen species. Indeed, methanimine might play a key role as an intermediate towards the formation of complex N-bearing hydrocarbons in the ionosphere that will lead, lower in the atmosphere, to the orange photochemical haze surrounding the moon.
   \begin{acronyms}
   \acro{EUV}
   Extreme UltraViolet
  \acro{APSIS}
   Atmospheric Photochemistry Simulated by Synchrotron
   \acro{INMS}
   Ion Neutral Mass Spectrometer 
    \acro{CAPS}
   CAssini Plasma Spectrometer
    \acro{IBS}
   Ion Beam Spectrometer
    \acro{ELS}
   ELectron Spectrometer
  \end{acronyms}

%

%
%
%
%
%
%
%
%

\acknowledgments
We thank Prof. G. Cernogora for his plasma expertise and fruitful discussions. This research was supported by the ERC Starting Grant PRIMCHEM, grant agreement 636829. According to the AGU Publications Data Policy, the data can be access without any restriction via the figshare general repository: \url{https://doi.org/10.6084/m9.figshare.9956399.v1}


%
%

\bibliography{agusample}

%
%
%
%
%

\end{document}


%
%


\title{Supporting Information for "Low Pressure EUV Photochemical Experiments: Insight on the Ion-Chemistry Occurring in Titan's Atmosphere"}
%
%

%
%



\authors{J. Bourgalais\affil{1}, N. Carrasco\affil{1}, L. Vettier\affil{1}, and P. Pernot\affil{2}}


\affiliation{1}{Universit\'e Versailles St-Quentin, Sorbonne Universit\'e, UPMC Univ. Paris 06, CNRS/INSU, LATMOS-IPSL, 11 boulevard d'Alembert, 78280 Guyancourt, France}
\affiliation{2}{Laboratoire de Chimie Physique, CNRS, Univ. Paris-Sud, Universit\'e Paris-Saclay, 91405, Orsay, France}

%
%

%

\begin{article}

%
%

\noindent\textbf{Additional Supporting Information (Files uploaded separately)}
\begin{enumerate}
\item Captions for Datasets S1 to Sx
\item Captions for large Tables S1 to Sx (if larger than 1 page, upload as separate excel file)
\item Captions for Movies S1 to Sx
\item Captions for Audio S1 to Sx
\end{enumerate}

\noindent\textbf{Introduction}


\noindent\textbf{Text S1.}
%


\noindent\textbf{Data Set S1.} 


\noindent\textbf{Movie S1.} 


\noindent\textbf{Audio S1.} 
\end{article}
\clearpage